\documentclass[twocolumn,showpacs,preprintnumbers,amsmath,amssymb]{revtex4}

\usepackage{graphicx}
\usepackage{bm}

\usepackage{amsfonts}

\newcommand{\erfc}{\mathop{\rm erfc}\nolimits}

\begin{document}
\title{Two-dimensional nonlocal vortices, multipole solitons and azimuthons
in dipolar Bose-Einstein condensates.}
\author{V. M. Lashkin}
\email{vlashkin@kinr.kiev.ua}
 \affiliation{Institute for Nuclear
Research, Pr. Nauki 47, Kiev 03680, Ukraine}

\date{\today}

\begin{abstract}
We have performed numerical analysis of the two-dimensional (2D)
soliton solutions in Bose-Einstein condensates with nonlocal
dipole-dipole interactions. For the modified 2D Gross–-Pitaevski
equation with nonlocal and attractive local terms, we have found
numerically different types of nonlinear localized structures such
as fundamental solitons, radially symmetric vortices, nonrotating
multisolitons (dipoles and quadrupoles), and rotating
multisolitons (azimuthons). By direct numerical simulations we
show that these structures can be made stable.
\end{abstract}


\maketitle

\section{Introduction}

The recent first experimental realization of a degenerate dipolar
atom gas \cite{Exper1}, where a Bose-Einstein condensate (BEC) of
$^{52}$Cr atoms has been observed, and optimistic perspectives in
creating a degenerate gas of polar molecules \cite{Exper2} have
stimulated a growing interest in the study of BEC with nonlocal
dipole-dipole interactions \cite{Santos05, Pedri06, Lushnikov}.
Dipole-dipole forces are anisotropic and long range, so that the
inter-particle interaction becomes essentially nonlocal.

Nonlocal nonlinearity naturally arises in many areas of nonlinear
physics and plays a crucial role in the dynamics of nonlinear
coherent structures. In particular, a rigorous proof of absence of
collapse in arbitrary spatial dimensions during the wave-packet
propagation described by the nonlocal nonlinear Schr\"{o}dinger
equation (NLSE) with sufficiently general symmetric response
kernel has been presented in Refs. \cite{Tur,Krol}. Stable vortex
\cite{Yakimenko,Briedis}, dipole \cite{We1,Lopez,Skupin,Kartashov}
and azimuthon \cite{Lopez,Skupin, We2} solitons in media with
nonlocal nonlinear response were theoretically predicted. Finally,
nonlocality induces attraction between solitons and allows for the
formation of bound states of out-of-phase bright solitons
\cite{Mironov} and dark solitons \cite{Nikolov}.

A very attractive feature of BEC with dipole-dipole interactions
is that the interplay between the nonlocal interaction, which is
only partially attractive and may be tuned by means of rotating
orienting fields \cite{Tune}, and the usual local short-range
contact forces, leads to the possibility of experimental
realization of highly controllable and stable solitary structures
in BEC \cite{Santos05}.

Recently, Pedro and Santos \cite{Santos05} have studied the
physics of bright solitons in two-dimensional (2D) dipolar
Bose-Einstein condensates with repulsive short-range
interactioins. Using the reduction procedure, they have obtained
2D modified Gross–-Pitaevski equation with the nonlocal term
describing dipole-dipole interaction and showed that the existence
of stable 2D solitary waves is possible.

In this paper, using 2D model suggested by Pedri and Santos
\cite{Santos05}, we study 2D solitary waves (SWs) in BEC with
attractive short-range and nonlocal dipole-dipole interactions. As
is known, the collapse of BEC's at some critical number of atoms
is the main consequence of the attractive nonlinearity
\cite{Dalfovo}. The presence of nonlocal interaction, however,
significantly changes the situation and leads to stable localized
states. We present different types of SWs (fundamental solitons,
vortices, nonrotating and rotating multisolitons) and by direct
numerical simulations show that these localized structures can be
made stable.

\section{Model and basic equations}
\label{sec2}

A dipolar BEC, consisting of $\mathcal{N}$ particles with the
dipole moment $d$ oriented along the $z$-axis, at sufficiently low
temperatures is described by a NLSE with nonlocal nonlinearity
\begin{gather}
i\hbar\frac{\partial \Psi}{\partial
t}=\left[-\frac{\hbar^{2}}{2m}\Delta+U(\mathbf{r})+g|\Psi|^{2}\right.
\nonumber \\
\left. +\int
V(\mathbf{r}-\mathbf{r'})|\Psi(\mathbf{r'})|^{2}\,d\mathbf{r'}\right]\Psi,
\end{gather}
where $\Psi(\mathbf{r},t)$ is the condensate wave function
normalized to the total number of particles: $\int
|\Psi(\mathbf{r})|^{2}\,d\mathbf{r}=\mathcal{N}$. The coupling
constant $g$ corresponds to the local contact interaction and
$g=4\pi\hbar^{2}a/m$, where $a$ is the $s$-wave scattering length.
In the following, we consider $a<0$, i.e. attractive short-range
interactions. An external trapping potential is assumed to be of
the form $U(\mathbf{r})=m\omega_{z}^{2}z^{2}/2$, with no trapping
in the $xy$-plane. All dipoles are assumed to be oriented along
the trap axis. The nonlocal potential is due to the dipole-dipole
interaction, and the kernel $V(\mathbf{r})$ is given by
$V(\mathbf{r})=g_{d}(1-3\cos^{2}\theta)/r^{3}$, where
$g_{d}=\alpha\mu_{0} d^{2}/4\pi$, $\theta$ is the angle between
the vector $\mathbf{r}$ and the dipole axis, $\mu_{0}$ is the
magnetic permeability of the vacuum, and $-1/2\leq\alpha\leq 1$ is
a tunable parameter \cite{Santos05, Tune}.

Assuming the anzatz
$\Psi(\mathbf{r})=\psi(\mathbf{r}_{\perp})\varphi_{0}(z)$, where
the function $\varphi_{0}(z)$, describing the condensate in the
direction of the tight confinement, is the ground state of the 1D
harmonic oscillator in the $z$-direction and normalizing the
length, time, and wave function by $l_{z}/\sqrt{2}$,
$1/\omega_{z}$, and $(\mathcal{N}/l_{z}^{3})^{1/2}$, respectively
(where $l_{z}=(\hbar/m\omega_{z})^{1/2}$), authors of Ref.
\cite{Santos05}, following the standard reduction procedure,
obtained the following 2D equation
\begin{equation}
\label{main2} i\frac{\partial\psi}{\partial
t}=-\Delta_{\perp}\psi+\bar{g}\psi\left\{|\psi|^{2} +\beta\int
R(\mathbf{r}-\mathbf{r}')|\psi(\mathbf{r}')|^{2}\,d\mathbf{r}'\right\},
\end{equation}
with the two free dimensionless parameters
\begin{equation}
\bar{g}=\frac{g}{\sqrt{2\pi}\hbar\omega_{z}l_{z}^{3}}=\frac{2\sqrt{2\pi}a}{l_{z}},\quad
\beta=\frac{g_{d}}{g}.
\end{equation}
The Fourier transform of the kernel $R(\mathbf{r})$ in Eq.
(\ref{main2}) is
\begin{equation}
\label{R}
 \hat{R}(\mathbf{k})=2-3\sqrt{\pi}k e^{k^{2}}\erfc(k),
\end{equation}
where $\erfc(x)$ is the complementary error function, so that
\begin{equation}
\label{R1} R(\mathbf{r})=\frac{1}{(2\pi)^{2}}\int
e^{i\mathbf{k}\cdot\mathbf{r}}\hat{R}(\mathbf{k})\,d\mathbf{k}.
\end{equation}
In what follows, since Eq. (\ref{main2}) admits an additional
rescaling, the parameter $\bar{g}$ has been fixed at $\bar{g}=\mp
1$, where the $-$($+$) sign corresponds to attractive (repulsive)
short-range interaction.

Equation (\ref{main2}) conserves the 2D norm, $N=\int
|\psi|^{2}dxdy$, and energy
\begin{gather}
E=\int \left\{\left|\frac{\partial\psi}{\partial
x}\right|^{2}+\left|\frac{\partial\psi}{\partial
y}\right|^{2}+\frac{1}{2}\bar{g}|\psi|^{4}+\frac{1}{2}\bar{g}\beta|\psi|^{2}\right.\nonumber
\\
\times\left.\left[\int
R(\mathbf{r}-\mathbf{r}')|\psi(\mathbf{r}')|^{2}\,d\mathbf{r}'\right]\right\}dxdy.
\end{gather}

\section{Modulational instability}

\begin{figure}
\includegraphics[width=3.4in]{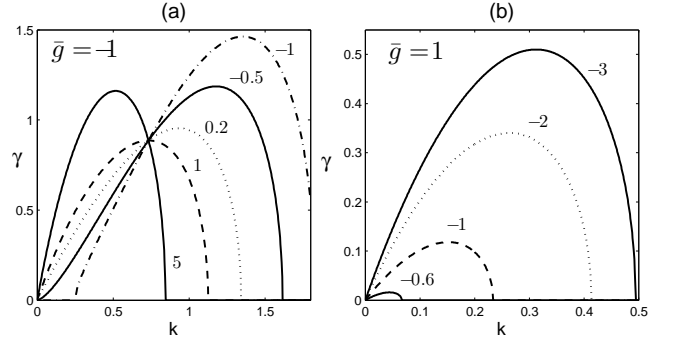}
\caption{\label{fig0} The growth rates $\gamma$ of the
modulational instability for (a) attractive ($\bar{g}=-1$) and (b)
repulsive ($\bar{g}=1$) short-range interactions. Values of
$\beta$ (the ratio between dipole-dipole and short-range
interaction) are shown near the curves. }
\end{figure}

An important feature of the dipole-dipole interaction is that, due
to the anisotropy, it is only partially attractive.
Correspondingly, the spectrum $\hat{R}(k)$ of the response
function $R(\mathbf{r})$ is not sign definite (note, in this
connection, that an analysis of 2D soliton dynamics in the
framework of Eq. (\ref{main2}) with somewhat resembling Eq.
(\ref{R}), but positive definite, kernel $\hat{R}(k)$ was
performed in Ref. \cite{Skupin}). Equation (\ref{main2}) has a
solution in the form of plane wave
\begin{equation}
\Psi_{0}=|\Psi_{0}|\exp(i\mathbf{k}_{0}\cdot\mathbf{r}-i\omega t)
\end{equation}
provided $\omega=k_{0}^{2}-\bar{g}(1+\beta\int
R(\mathbf{r})d\mathbf{r})|\Psi_{0}|^{2}$. The stability properties
of the plane wave essentially depend on sign definiteness of the
spectrum of nonlinear response function \cite{Krol,Wyller}. On the
other hand, modulational instability (MI) (instability of the
plane wave with amplification of both sidebands) is often
considered as a precursor for the formation of bright solitons.
Considering perturbed plane wave solutions in the form
\begin{equation}
\psi=(|\Psi_{0}|+\delta\psi)\exp(i\mathbf{k}_{0}\cdot\mathbf{r}-i\omega
t),
\end{equation}
where
\begin{equation}
\delta\psi=\psi_{+}e^{i\mathbf{k}\cdot\mathbf{r}-\gamma
t}+\psi_{-}e^{-i\mathbf{k}\cdot\mathbf{r}+\gamma t} ,
\end{equation}
and linearizing Eq. (\ref{main2}) around $\Psi_{0}$ in
$\delta\psi$, one can obtain the growth rate $\gamma$ of MI of
homogeneous field ($k_{0}=0$) for the model Eq. (\ref{main2})
\begin{equation}
\label{gam}
\gamma^{2}=-2|\Psi_{0}|^{2}k^{2}\bar{g}[1+\beta\hat{R}(k)]-k^{4}.
\end{equation}
Instability occurs if $\gamma^{2}>0$. In Figure ~\ref{fig0} we
show the dependence of the growth rate of MI on $k$ for the cases
of attractive ($\bar{g}=-1$) and repulsive ($\bar{g}=1$)
short-range interactions. In the attractive case, the growth rate
is equal to zero for $0\leq k<k_{cr}$, where $k_{cr}$ is some
critical value depending on $\beta$ (the ratio between
dipole-dipole and short-range interaction), if $\beta\lesssim 0.5$
(i. e., in particular, for all negative $\beta$), so that long
wave modes are stable. Optimal, i.e. corresponding to maximum of
the growth rate, wave number $k_{opt}$ decreases with increasing
$\beta$. In the repulsive case, the growth rate of MI is equal to
zero for all positive $\beta$ and for $\beta  \gtrsim -0.4$. This
is in agreement with results of Ref. \cite{Santos05}, where bright
solitons (for the repulsive case $\bar{g}=1$) were predicted only
for negative $\beta$ and $|\beta|>0.12$.

\section{Numerical results}

\label{sec3}

\begin{figure*}
\begin{center}\includegraphics[width=6.8in]{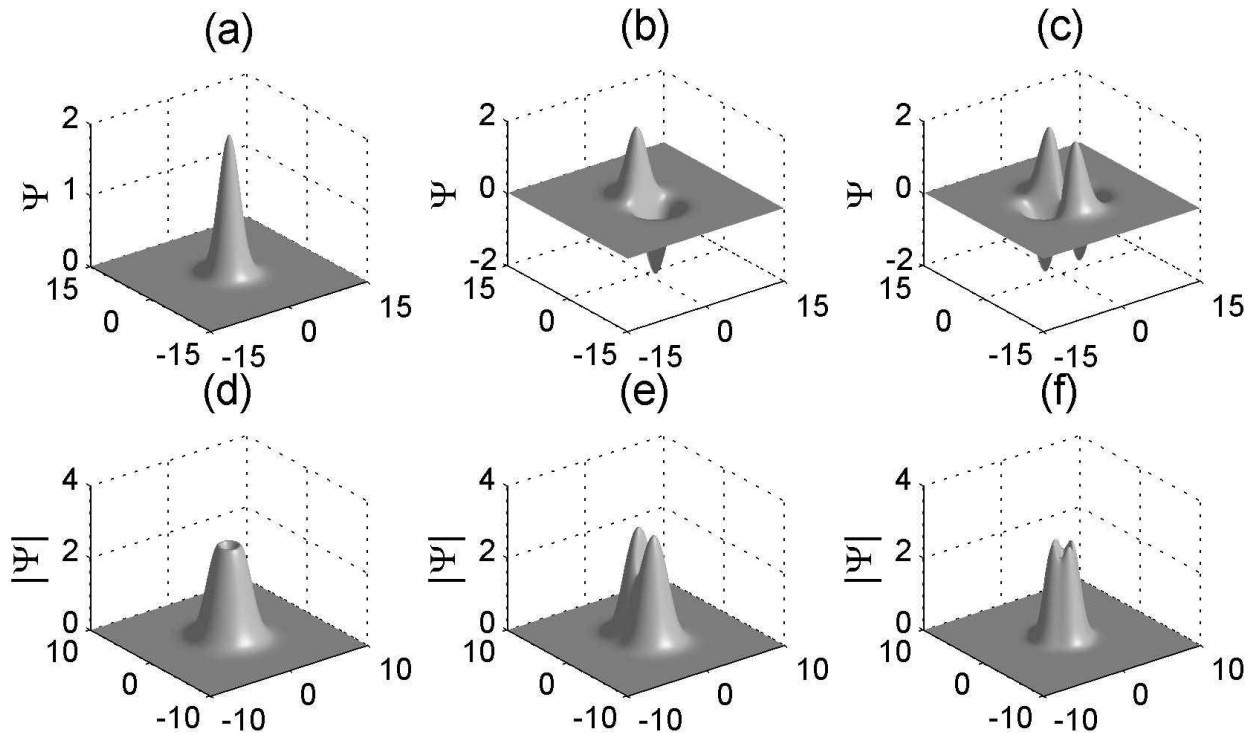}\end{center}
\caption{\label{fig1} Numerically found stationary localized
nonrotating (a-c) and rotating (d-f) solutions of Eq.
(\ref{main2}) with $\beta=2$: (a) monopole with $\mu=-2$; (b)
dipole with $\mu=-2$; (c) quadrupole with $\mu=-2$; (d) vortex
with $\mu=-5$; (e) azimuthon with $\mu=-5$, $p=0.6$ and two
intensity peaks; (f) azimuthon with $\mu=-5$, $p=0.9$ and four
intensity peaks. }
\end{figure*}

We look for stationary solutions of Eq. (\ref{main2}) with
$\bar{g}=-1$ (attractive short-range interaction) in the form
$\psi(x,y,t)=\Psi(x,y)\exp(-i\mu t)$, where $\mu$ is the chemical
potential, so that $\Psi$ obeys the equation
\begin{equation}
\label{main3} (\mu+\Delta_{\perp})\Psi=-\Psi\left\{|\Psi|^{2}
+\beta\int
R(\mathbf{r}-\mathbf{r}')|\Psi(\mathbf{r}')|^{2}\,d\mathbf{r}'\right\},
\end{equation}
where $R(\mathbf{r})$ is determined by Eqs. (\ref{R}) and
(\ref{R1}). To solve numerically Eq. (\ref{main3}), we impose
periodic boundary conditions on Cartesian grid and use the
relaxation technique similar to one described in Ref.
\cite{Petviashvili}. We have not found any localized solutions
with $\mu>0$. Fundamental soliton solutions of Eq. (\ref{main3})
with $\beta<0$ or $\beta>\beta_{cr}$, where $\beta_{cr}\sim 2.1$,
turn out to be unstable with $\partial N/\partial\mu>0$. Thus, in
what follows, we consider the region $0<\beta<\beta_{cr}$ and
specifically set $\beta=2$. Choosing an appropriate initial guess,
one can find numerically with high accuracy (the norms of the
residuals were less than $10^{-9}$) three different classes of
spatially localized solutions of Eqs. (\ref{main3}) -- the
nonrotating (multi)solitons, the radially symmetric vortices, and
the rotating multisolitons (azimuthons).

The real (or containing only a constant complex factor) function
$\Psi(x,y)$ corresponds to nonrotating solitary structures.
Examples of such nonrotating (multi)solitons for Eq.
(\ref{main3}), namely, a monopole, a dipole, and a quadrupole are
presented in Figs.~\ref{fig1}(a)-~\ref{fig1}(c). The nonrotating
multipoles consist of several fundamental solitons (monopoles)
with opposite phases.

The second class of solutions, vortex solutions, are the solutions
with the radially symmetric amplitude $|\Psi(x,y)|$, that vanishes
at the center, and a rotating spiral phase in the form of a linear
function of the polar angle $\theta$, i.e. $\arg\Psi=m\theta$,
where $m$ is an integer. The index $m$  (topological charge)
stands for a phase twist around the intensity ring. The important
integral of motion associated with this type of solitary wave is
the angular momentum, which can be expressed through the vortex
amplitude and phase. The numerically found single-charged ($m=1$)
vortex solution of Eq. (\ref{main3}) is shown in
Fig.~\ref{fig1}(d).

\begin{figure}
\includegraphics[width=3.4in]{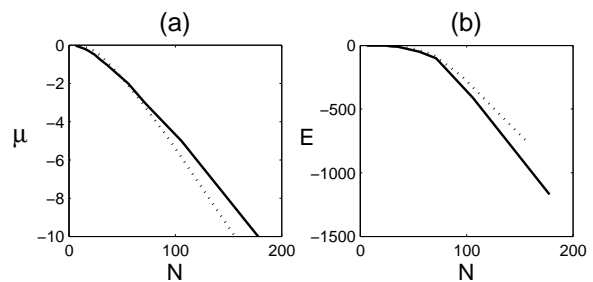}
\caption{\label{fig2} (a) Chemical potential $\mu$ and (b) energy
$E$ versus normalized number of atoms $N$. Solid line: nonrotating
dipoles ($p=0$). Dotted line: vortices ($p=1$). }
\end{figure}

The third class of solutions, rotating multisolitons with the
spatially modulated phase, were first introduced in Ref.
\cite{Kivshar1} for models with local nonlinearity, where they
were called azimuthons. The azimuthons can be viewed as an
intermediate kind of solutions between the rotating radially
symmetric vortices and nonrotating multisolitons. Using
variational analysis to describe azimuthons, the authors of Ref.
\cite{Lopez} considered the following trial function in polar
coordinates ($r$,$\theta$)
\begin{equation}
\label{trial} \Psi(r,\theta)=r^{|m|}\Phi(r)(\cos m\theta+ip\,\sin
m\theta),
\end{equation}
where $\Phi$ is the real function, which vanish fast enough at
infinity, $m$ is an integer, and $0\leq p\leq 1$. The case $p=0$
corresponds to the nonrotating multisolitons (e. g. $m=1$ to a
dipole, $m=2$ to a quadrupole etc.), while the opposite case $p=1$
corresponds to the radially symmetric vortices. The intermediate
case $0<p<1$ corresponds to the azimuthons. In our case, the
numerically found complex function $\Psi(x,y)$ with a spatially
modulated phase corresponds to the azimuthons. We introduced the
parameter $p$ (modulational depth), which is similar to the one in
Eq. (\ref{trial}), in the following way
\begin{equation}
p=\max|\mathrm{Im}\,\Psi|/\max|\mathrm{Re}\,\Psi|.
\end{equation}
For fixed chemical potential $\mu$, there is a family of
azimuthons with different $p$. Like the radially symmetric
vortices, the azimuthons carry out the nonzero angular momentum.
In Figures~\ref{fig1}(e) and ~\ref{fig1}(f) we demonstrate two
numerically found examples of the azimuthons for the nonlocal
model described by Eqs. (\ref{main2}). Figure ~\ref{fig2} shows
the dependences of the chemical potential $\mu$ and energy $E$ on
the normalized number of atoms $N$, for the dipoles ($p=0$) and
vortices ($p=1$).

We next addressed the stability of these localized solutions and
study the evolution of the solitons in the presence of small
initial perturbations. We have undertaken extensive numerical
modeling of Eq. (\ref{main2}) initialized with our computed
solutions with added gaussian noise. The initial condition was
taken in the form $\Psi(x,y)[1+\varepsilon \Phi(x,y)]$, where
$\Psi(x,y)$ is the numerically calculated exact solution,
$\Phi(x,y)$ is the white gaussian noise with variance
$\sigma^{2}=1$ and the parameter of perturbation
$\varepsilon=0.005 \div 0.1$. In addition, azimuthal perturbation
of the form $i\varepsilon\sin\theta$ was taken for the vortices
and azimuthons. Spatial discretization was based on the
pseudospectral method. Under this, since the Fourier transform of
the kernel $R$ is known, the nonlocal term can be easily computed
with the aid of the convolution theorem. Temporal
$t$-discretization included the split-step scheme. As was said
above, we consider the region $0<\beta<\beta_{cr}$.

The fundamental solitons are stable for all $\mu<0$. These
solitons have $\partial N/\partial\mu <0$ so that the
Vakhitov-Kolokolov stability criterion is met.

\begin{figure}
\includegraphics[width=3.4in]{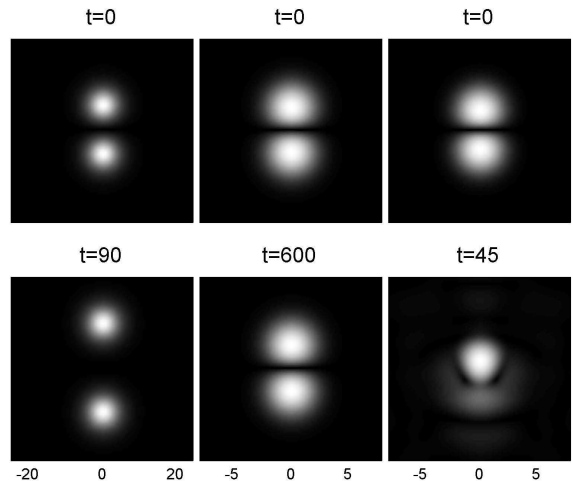}
\caption{\label{fig3} Left column: Splitting of the dipole with
$\mu=-0.2$ into two monopoles; middle column: stable dynamics of
the dipole with $\mu=-3$; right column: unstable dipole with
$\mu=-4$. }
\end{figure}

 Depending
on the parameter $\mu$, we observed three different scenarios of
the nonrotating dipole evolution, which are presented in
Fig.~\ref{fig3} (for $\varepsilon=0.01$). The first regime
corresponds to the region $\mu_{cr}<\mu<0$, and for $\beta=2$ we
found $\mu_{cr}\sim -0.2$, which corresponds to the normalized
number of atoms $N_{cr}\sim 15.5$. If $\mu_{cr}<\mu$, the initial
dipole splits in two monopoles which move in the opposite
directions without changing their shape and without radiation,
i.~e. the monopoles just go away at infinity. This type of the
evolution is shown in the left column of Fig.~\ref{fig3}. Under
this, the value $\delta N=N_{dip}-2N_{mon}$, where $N_{dip}$ and
$N_{mon}$ are 2D norms for the dipole and monopole respectively,
tends to almost zero as $\mu$ approaches $\mu_{cr}$.

The second regime of the dipole evolution corresponds to the
region $\mu_{th}<\mu<\mu_{cr}$, where (for $\beta=2$)
$\mu_{th}\sim -3.1$. The numerical simulations clearly show that
in this range of the parameter $\mu$ the dipoles are stable with
respect to initial noisy perturbations and survive over huge
times. In terms of the 2D norm (normalized number of atoms), the
stability region is written as $N_{cr}<N<N_{th}$, where
$N_{th}\sim 71$. The stable dynamics of the dipole is illustrated
in the middle column of Fig.~\ref{fig3} (for $\beta=2$, $\mu=-3$
and $\varepsilon=0.01$).

The further (after $\mu_{th}\sim -3.1$) decreasing of the chemical
potential $\mu$ (or, equivalently, increasing of the normalized
number of atoms $N$) sharply shortens the times at which the
dipole survives, and, the dipoles with $\mu<\mu_{th}$ are
unstable. The typical decay of the unstable dipole below the
threshold value $\mu_{th}$ of the chemical potential is shown in
the right column of Fig.~\ref{fig3}. Thus, the stable dipoles
exist only within a finite, rather narrow range of the normalized
number of atoms $N$.

\begin{figure}
\includegraphics[width=3.4in]{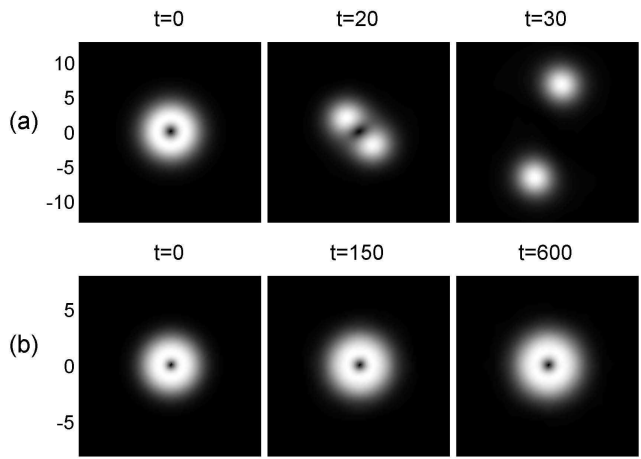}
\caption{\label{fig4} (a) Splitting of the vortex with $\mu=-1$;
(b) stable dynamics of the vortex with $\mu=-5$. }
\end{figure}

A somewhat different behavior we observed for the vortices. The
numerical simulations clearly show that the vortices with
$\mu<\mu_{cr}$, where $\mu_{cr}$ is some critical value and for
$\beta=2$ we found $\mu_{cr}\sim -1.4$ (with corresponding
$N_{cr}\sim 45$), are stable with respect to small initial noisy
and azimuthal perturbations up to the maximum times used (of the
order of $t=1000$). The vortices with $\mu>\mu_{cr}$ (i.e.
$N<N_{cr}$) splits in two fundamental solitons moving in the
opposite directions. These two different scenarios of the vortex
evolution are illustrated in Fig.~\ref{fig4}. Thus, the vortices
can be made stable if the 2D norm (normalized number of atoms)
exceeds some critical value $N_{cr}$.

\begin{figure}
\includegraphics[width=3.4in]{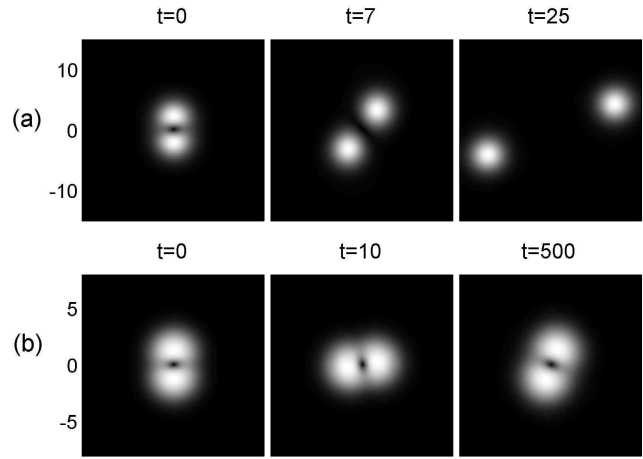}
\caption{\label{fig5} (a) Splitting of the azimuthon with two
intensity peaks and $\mu=-1$, $p=0.4$; (b) stable dynamics of the
azimuthon with two intensity peaks and $\mu=-5$, $p=0.6$. }
\end{figure}

We have not performed numerical analysis of the azimuthon
evolution for different $\lambda$ and arbitrary $p$ because of the
difficulties in finding azimuthon solutions with arbitrary $p$.
Nevertheless, we can conclude that azimuthons with two intensity
peaks and not too small $p$ can be made stable if the 2D norm
(normalized number of atoms) $N$ exceeds some critical value
depending on $p$. Splitting of the azimuthon with two intensity
peaks and $\mu=-1$, $p=0.4$, and stable dynamics of the azimuthon
with $\mu=-5$ and $p=0.6$ are shown in Fig.~\ref{fig5}.
Numerically estimated rotational velocity of the stable azimuthon
in Fig.~\ref{fig5}(b) is $\omega=0.18$ so that it survives over
many dozens of rotational periods.

Note, that one point should be emphasized. Strictly speaking, our
direct numerical modelling  can not  give a rigorous proof of
stability/instability of the multisolitons. First, in the the
direct numerical experiments one can consider the evolution over
finite times only. Second, the results are limited to the
perturbation profile. A rigorous proof could, for instance,
include a linear stability analysis with the corresponding
eigenvalue problem. Nevertheless, from our numerical simulations
of the dynamics over finite, but large, times we can conclude that
(in stable cases) the potential growth rates of unstable modes are
very small. The structures (if stable) survive over huge times and
hundreds of rotational periods, and from the practical point of
view they can be regarded as stable.

\section{Conclusion}

In conclusion, we have demonstrated the existence of 2D localized
nonlinear structures in BECs with nonlocal dipole-dipole and
attractive short-range contact interactions and studied their
stability. We have found numerically three kinds of soliton
families: nonrotating multipole solitons (fundamental one-hump
soliton, dipole and quadrupole), radially symmetric vortices, and
rotating multihump (with two and four intensity peaks) solitons
with the spatially modulated phase (azimuthons). We have shown
that stable solitons may exist only within a finite range of the
ratio between dipole-dipole and short-range interactions (both of
which are tunable). The anisotropy of the dipole-dipole
interaction is crucial, since this leads to partially attractive
nature of the interaction. Sufficiently large dipolar interactions
destabilize the SWs. By direct numerical simulations, we have
found that dipole nonrotating solitons, vortices and two intensity
peak azimuthons can be stable for some values of the chemical
potential (or, equivalently, normalized number of atoms).

\begin{acknowledgments}
The author is grateful to A. I. Yakimenko and Yu. A. Zaliznyak for
useful discussions.
\end{acknowledgments}

\end{document}